\documentstyle[preprint,eqsecnum,aps,epsfig,axodraw]{revtex}
\tightenlines
\begin{document}
\draft
\title{
\rightline{\rm \normalsize INFNCA-TH-0010}
Neutrino self-energy in a magnetized medium in arbitrary $\xi$-gauge}
\author{A. Erdas \thanks{Email address: andrea.erdas@ca.infn.it}}
\address{
Dipartimento di Fisica dell' Universit\`a di Cagliari
and I.N.F.N. Sezione di Cagliari,
\\
Cittadella Universitaria, S.P. per Sestu Km 0.700,
I-09042 Monserrato (CA), Italy
}
\author{C. Isola \thanks{Email address: claudia.isola@cpht.polytechnique.fr}}
\address{Centre de Physique Theorique,
\\
Ecole Polytechnique, 91128 Palaiseau Cedex, France}
\maketitle
\begin {abstract} 
We calculate the one-loop neutrino self-energy in a magnetized plasma to all orders
in the magnetic field. The calculation is done in a general gauge. We obtain the
dispersion relation and effective potential for neutrinos in a $CP$-symmetric plasma
under various conditions, and show that, while the self-energy depends on the gauge 
parameter $\xi$, the dispersion relation and effective potential to leading order
are independent of it.
\end {abstract}
%%%%%%%%%%%%%%%%%%%%%%%%%%%%%%%%%%%%%%%%%%%%%%%%%%%%%%%%%%%%%%%%%%%%%
\section{Introduction}

The properties of neutrinos that propagate through a magnetized medium differ significantly
from those in vacuum: the vacuum energy-momentum dispersion relation is not valid in a
magnetized medium, the neutrino index of refraction becomes anisotropic and so does
the effective potential.

In the present work we wish to investigate the self-energy and thermal dispersion relation
for neutrinos in a medium with magnetic field.

The dispersion of neutrinos in a medium has been widely investigated in the literature:
in the presence of a $CP$-asymmetric background the thermal self-energy is proportional to 
the particle-antiparticle asymmetry and is manifestly gauge independent at leading order,
and the same is true for the leading correction to the vacuum dispersion relation.
The situation is different in a $CP$-symmetric medium, where
the next order term must be considered, since the 
particle-antiparticle asymmetry vanishes, and it occurs that to this order the 
thermal self-energy depends on the gauge parameter, while 
the dispersion relation is still gauge invariant, as shown by D' Olivo
et al. \cite{dolivo1}.

The vacuum dispersion relation in the presence of a constant magnetic field has also 
been studied \cite{erdasfeld}, and both the self-energy and dispersion relation are
gauge independent to leading order.

The neutrino dispersion relation in a magnetized medium at finite temperature
has been calculated before,
for $CP$-asymmetric and $CP$-symmetric background,
but in all these papers \cite{dolivo2,elmfors1,erdas2} the calculations are carried out 
in a particular gauge and therefore
it is not possible to investigate the gauge dependence of the results.

In this paper we do a detailed calculation of the exact neutrino self-energy in a
magnetized background,
the calculation is carried out in an arbitrary gauge $\xi$ and we use the real time method
of finite temperature field theory. Next we consider a $CP$-symmetric plasma and 
show that, although the self-energy depends
on the gauge parameter, the dispersion relation and effective potential are independent
of $\xi$ up to leading order terms.

In Sec. II we calculate the thermal self-energy to all orders in the magnetic field
in an arbitrary gauge. In Sec. III we calculate the thermal self-energy in arbitrary 
gauge in a $CP$-symmetric magnetized medium. First we consider a temperature $T$ 
much smaller than 
the mass $M$ of the $W$-boson and we show explicitly that the dispersion relation 
and the effective potential
do not depend on the gauge parameter up to terms of order $(g^2/M^4)$. 
Then we calculate the self-energy for $T\sim M$
and show that also in this case the dispersion relation is independent of the gauge
parameter to leading order. In Sec. IV we draw our conclusions.
%%%%%%%%%%%%%%%%%%%%%%%%%%%%%%%%%%%%%%%%%%%%%%%%%%%%%%%%%%%%%%%%%%%%%
\section{ Thermal self-energy in a constant
magnetic field in the $\xi$-gauge}
In this section we calculate the real part of the 
neutrino thermal self-energy in a magnetized medium using the exact fermion, scalar and 
gauge boson propagators in a constant magnetic field in an arbitrary gauge $\xi$. We take the
magnetic field pointing along the positive $z$-direction and indicate with $B$ its magnitude,
therefore the only non-vanishing components of the electromagnetic 
field strength tensor $F^{\mu \nu}$ are $F^{12}=B=-F^{21}$. The exact
expression for the electron 
$S_0(x',x'')$ \cite{schwinger,dittrich}, $W$ boson 
$G_0^{\mu \nu}(x',x'')$ \cite{erdasfeld} and charged scalar
$\Delta_0(x',x'')$ \cite{erdasfeld,dittrich} vacuum propagators in a constant magnetic field
are obtained using Schwinger's proper time method
\begin{equation}
S_0(x',x'')=\phi^\ast(x',x'')\!\int{d^4k\over (2\pi)^4}e^{ik\cdot (x'-x'')}
S_0(k),
\label{2_01a}
\end{equation}
\begin{equation}
G_0^{\mu\nu}(x',x'')=\phi(x',x'')\!\int{d^4k\over (2\pi)^4}e^{ik\cdot (x'-x'')}
G_0^{\mu\nu}(k)
\label{2_02a}
\end{equation}
\begin{equation}
\Delta_0(x',x'')=\phi(x',x'')\!\int{d^4k\over (2\pi)^4}e^{ik\cdot (x'-x'')}
\Delta_0(k)
\label{2_03a}
\end{equation}
where the translationally invariant parts of the propagators are
\begin{equation}
S_0(k)=i
\int_0^\infty \!\!{ds\over\cos eBs}\,
{\exp{\left[-is\left(m^2-i\epsilon+k^2_{\scriptscriptstyle{\parallel}}+
k^2_{\scriptscriptstyle{\perp}}{\tan eBs\over eBs}\right)\right]}}
\left[(m-
\not\! k_{\scriptscriptstyle{\parallel}})e^{-ieBs\sigma_3}-
{\not\! k_{\scriptscriptstyle{\perp}}\over \cos eBs}
\right],
\label{2_01}
\end{equation}
\begin{eqnarray}
G_0^{\mu \nu}(k)&=&i
\int_0^\infty \!\!{ds\over\cos eBs}\,
{\exp{\left[-is\left(k^2_{\scriptscriptstyle{\parallel}}+
k^2_{\scriptscriptstyle{\perp}}{\tan eBs\over eBs}\right)\right]}}
\Biggl\{
e^{-is(M^2-i\epsilon)}[g^{\mu \nu}_{\scriptscriptstyle{\parallel}}
+(e^{2eFs})^{\mu\nu}_{{\scriptscriptstyle{\perp}}}]
\nonumber \\
&&+
\left[\left(k^{\mu}+k_{\lambda}
F^{\mu \lambda}{\tan eBs\over B}\right)
\left(k^{\nu}+k_{\rho}
F^{\rho \nu}{\tan eBs\over B}\right)
+i{e\over 2}\left(F^{\mu\nu}-
g^{\mu\nu}_{{\scriptscriptstyle{\perp}}}B\tan eBs
\right)\right]
\nonumber \\
&&\times
\left({e^{-is(M^2-i\epsilon)}-e^{-is(\xi M^2-i\epsilon)}\over M^2}\right)
\Biggr\},
\label{2_02}
\end{eqnarray}
\begin{equation}
\Delta_0(k)=i\int_0^\infty\!\!\!{ds\over\cos eBs}\,
{\exp{\left[-is\left(\xi M^2-i\epsilon+k^2_{\scriptscriptstyle{\parallel}}+
k^2_{\scriptscriptstyle{\perp}}{\tan eBs\over eBs}\right)\right]}}
\label{2_03}
\end{equation}
We have used the subscript $ _{0}$ on the three propagators to indicate that
these are still the vacuum propagators,
$-e$ and $m$ are the charge  and mass of the electron, 
$M$ the $W$-mass, $\xi$ the gauge parameter, $\sigma_3=\sigma^{12}=
{i \over 2}[\gamma^1, \gamma^2]$, and the metric we use is
$g^{\mu \nu} = {\rm diag}(-1,+1,+1,+1)$.
We choose the electromagnetic vector potential to be
$A_\mu=-{1\over2}F_{\mu \nu}x^\nu$ and therefore
the phase factor in Eqs. (\ref{2_01a}), (\ref{2_02a}), and (\ref{2_03a}) is
\cite{dittrich}
\begin{equation}
\phi(x',x'')=\exp\left[
ie\int^{x'}_{x''}dx_\mu A^\mu(x)
\right]=\exp\left(
i {e\over 2}x''_\mu F^{\mu \nu} x'_\nu
\right)
\label{2_07}
\end{equation}
because the integral is independent of the choice of integration path.
For any 4-vector $a^\mu$ we use the notation \cite{erdasfeld,erdas2,dittrich}
$a^\mu_{\scriptscriptstyle{\parallel}}=(a^0,0,0,a^3)$ and
$a^\mu_{\scriptscriptstyle{\perp}}=(0,a^1,a^2,0)$
and we have
\begin{equation}
\left(e^{2eFs}\right)^{{\mu}\nu}_{{\scriptscriptstyle{\perp}}}=
(g^{\mu\nu})_{\scriptscriptstyle{\perp}} \cos 2eBs 
+F^{\mu\nu} {\sin 2eBs\over B}\,\,.
\end{equation}

The expressions for  the real-time thermal propagators in a magnetized medium are easily 
obtained using the vacuum propagators \cite{elmfors1,elmfors2}. The electron thermal propagator is
\begin{equation}
S(x',x'')=S_0(x',x'')-\phi^\ast(x',x'')\!\int{d^4k\over (2\pi)^4}e^{ik\cdot (x'-x'')}
[S_0(k)-S_0^*(k)]f_F(k^0),
\label{2_09}
\end{equation}
the $W$-boson propagator is
\begin{equation}
G^{\mu\nu}(x',x'')=G_0^{\mu\nu}(x',x'')+\phi(x',x'')\!\int{d^4k\over (2\pi)^4}e^{ik\cdot (x'-x'')}
[G_0^{\mu\nu}(k)-G_0^{\mu\nu *}(k)]f_B(k^0)
\label{2_10}
\end{equation}
and the scalar propagator is
\begin{equation}
\Delta(x',x'')=\Delta_0(x',x'')+\phi(x',x'')\!\int{d^4k\over (2\pi)^4}e^{ik\cdot (x'-x'')}
[\Delta_0(k)-\Delta_0^*(k)]f_B(k^0),
\label{2_11}
\end{equation}
where the fermion and boson occupation numbers are 
\begin{equation}
f_F(k^0)=f_F^+(k^0)\theta(k^0)+f_F^-(k^0)\theta(-k^0)
\label{2_12}
\end{equation}
with
\begin{equation}
f_F^\pm(k^0)={1\over e^{\pm(k^0-\mu)/T}+1}
\label{2_13}
\end{equation}
and
\begin{equation}
f_B(k^0)={1\over e^{|k^0|/T}-1}
\end{equation}
$T$ is the temperature of the medium and $\mu$ the fermion chemical potential.

In the remainder of this paper we focus our attention on electron-type neutrinos,
the generalization to $\mu$ and $\tau$-neutrino is straightforward.
In order to calculate the one-loop electron neutrino self-energy in a magnetized medium, 
one needs to consider three Feynman diagrams, the tadpole $\Sigma_{tad}$,
the bubble diagram with a $W$ boson 
$\Sigma_{bos}(x',x'')$, and the bubble diagram with
a scalar $\Sigma_{scal}(x',x'')$, the first one gives only a local contribution to the
self-energy, while the last two give non-local contributions. They are shown in Fig. 1.
We are only interested in the real part of the self-energy, and in this case it is
sufficient to calculate only the 1-1 element of the self-energy in the real-time
formalism of finite temperature field theory and then take its real part.
The tadpole diagram does not depend on the gauge parameter $\xi$ because the $Z$ boson does
not carry any momentum, and so we use the result from the literature \cite{elmfors1}. 
In the case of an electron-rich medium
\begin{equation}
\Sigma_{tad}={g^2_Z\over
M_Z^2}\gamma_R\left[\left(-{1\over 2}+2\sin^2\theta_W\right)\gamma^0(N_e-N_{\bar{e}})
+{1\over 2} \gamma^3 (N_e^0-N_{\bar{e}}^0)
\right]\gamma_L ,
\label{2_15}
\end{equation}
where
\begin{equation}
g_Z={g\over 2\cos\theta_W}
\,\,,\gamma_R={1+\gamma_5\over2}\,\,\,,\,\,\,
\gamma_L={1-\gamma_5\over2}.
\label{2_16}
\end{equation}
The net number density of electrons in a magnetic field appears in Eq. (\ref{2_15}) and is 
defined as \cite{elmfors1}
\begin{equation}
N_e-N_{\bar{e}}={|eB| \over 2\pi^2}\int_0^{\infty}dk_z
\sum_{n=0}^{\infty}\sum_{\lambda=\pm1}\left[f_F^+(E_{n,\lambda,k_z})
-f_F^-(-E_{n,\lambda,k_z})\right]
\label{2_17}
\end{equation}
where $n$ is the orbital quantum number, $\lambda$ the spin
quantum number and the energies of the Landau levels are given by
\begin{equation}
E_{n,\lambda,k_z}^2=m^2+k^2_z+|eB|(2n+1-\lambda).
\label{2_18}
\end{equation}
The net number density in the lowest Landau level also appears in Eq.
(\ref{2_15}) and is given by
\begin{equation}
N_e^0-N_{\bar{e}}^0={|eB| \over 2\pi^2}\int_0^{\infty}dk_z
\left[f_F^+(E_{0,1,k_z})-f_F^-(-E_{0,1,k_z})\right]\,\,.
\label{2_19}
\end{equation}
Now we proceed to calculate the two bubble diagrams in an arbitrary gauge.
The two propagators in the loop belong to oppositely charged particles, 
and therefore the phase factors $\phi(x',x'')$ cancel out, giving a result which
is translationally invariant
\begin{equation}
\Sigma_{i}(x',x'')=\int{d^4p\over (2\pi)^4}e^{ip\cdot (x'-x'')}
\Sigma_{i}(p).
\label{2_20}
\end{equation}
where $i=bos\, ,\, scal$ . We can write
\begin{equation}
\Sigma_i(p)=\Sigma_i^0(p)+
\Sigma_i^F(p)+\Sigma_i^B(p)\,\,,
\label{2_21}
\end{equation}
where $\Sigma_i^0(p)$ is the vacuum part \cite{erdasfeld}, $\Sigma_i^F(p)$
the contribution of thermal electrons and $\Sigma_i^B(p)$ the contribution
of thermal bosons. The thermal parts are known in the $\xi=1$ gauge
\cite{erdas2}. The scalar contributions 
in an arbitrary gauge are immediately obtained by taking $\Sigma_{scal}^F(p)$
and $\Sigma_{scal}^B(p)$ in the $\xi=1$ gauge \cite{erdas2} and replacing 
$M^2$ with $\xi M^2$. The $W$-boson contribution is known in the $\xi=1$
gauge, and therefore we only need to calculate 
$\Sigma_{bos}(p,\xi)-\Sigma_{bos}(p,\xi=1)$ which is the difference
between the thermal part of the bubble diagram in the $\xi$-gauge and
the thermal part in the $\xi=1$ gauge. We have
\begin{eqnarray}
 &&\Sigma_{bos}^F(p,\xi)=
 \Sigma_{bos}^F(p,\xi\!\! =1 \!)+
 i{g^2\over 2M^2}\gamma_R\gamma_\mu \int {d^4k\over
(2\pi)^4}f_F(k^0)\int_{-\infty}^{+\infty}ds
{e^{-[is(m^2+\tilde{k}^2)+|s|\epsilon]} \over \cos eBs}
\nonumber \\
&&\times
\left[(m-
\not\! k_{\scriptscriptstyle{\parallel}})e^{-ieBs\sigma_3}-
{\not\! k_{\scriptscriptstyle{\perp}}\over \cos eBs}
\right]
\int_0^{\infty}dt
{e^{-it(\tilde{q}^2-i\epsilon)} \over \cos eBt}
\left(e^{-itM^2}-e^{-it\xi M^2}\right)
\nonumber \\
&&\times
\left[\left(q^{\mu}+q_{\lambda}
F^{\mu \lambda}{\tan eBt\over B}\right)
\left(q^{\nu}+q_{\rho}
F^{\rho \nu}{\tan eBt\over B}\right)
+i{e\over 2}\left(F^{\mu\nu}-
g^{\mu\nu}_{{\scriptscriptstyle{\perp}}}B\tan eBt
\right)\right]
\gamma_\nu\gamma_L ,
\label{2_22}
\end{eqnarray}
\begin{eqnarray}
 &&\Sigma_{bos}^B(p,\xi)=
 \Sigma_{bos}^B(p,\xi\!\! =1 \!)
 -i{g^2\over 2M^2}\gamma_R\gamma_\mu \int {d^4k\over
(2\pi)^4}f_B(k^0)\int_{0}^{\infty}ds
{e^{-is(m^2+\tilde{q}^2-i\epsilon)} \over \cos eBs}
\nonumber \\
&&\times
\left[(m-
\not\! q_{\scriptscriptstyle{\parallel}})e^{-ieBs\sigma_3}-
{\not\! q_{\scriptscriptstyle{\perp}}\over \cos eBs}
\right]
\int_{-\infty}^{+\infty}dt
{e^{-(it\tilde{k}^2+|t|\epsilon)} \over \cos eBt}
\left(e^{-itM^2}-e^{-it\xi M^2}\right)
\nonumber \\
&&\times
\left[\left(k^{\mu}+k_{\lambda}
F^{\mu \lambda}{\tan eBt\over B}\right)
\left(k^{\nu}+k_{\rho}
F^{\rho \nu}{\tan eBt\over B}\right)+i{e\over 2}\left(F^{\mu\nu}-
g^{\mu\nu}_{{\scriptscriptstyle{\perp}}}B\tan eBt
\right)\right]
\gamma_\nu\gamma_L ,
\label{2_23}
\end{eqnarray}
where $q=p-k$,
and we use the notation
\begin{equation}
s\tilde{k}^2=sk^2_{\scriptscriptstyle{\parallel}}+
k^2_{\scriptscriptstyle{\perp}}{\tan eBs\over eB}\;,\;\;\;\;\;
\;\;\;\;\;\;\;
t\tilde{k}^2=tk^2_{\scriptscriptstyle{\parallel}}+
k^2_{\scriptscriptstyle{\perp}}{\tan eBt\over eB}\,\,,\,\,
\label{2_24}
\end{equation}
and the same for $\tilde{q}$.
After some $\gamma$-algebra, integration over the $\vec k$-variable,
and change of the $t$ and $s$ variables, we obtain

\begin{eqnarray}
&&\Sigma_{bos}^F(p,\xi)
= \Sigma_{bos}^F(p,\xi\!\! =1 \!)-
{g^2\over 4M^2}\gamma_R\int_{-\infty}^{\infty}{dk^0\over (2\pi)^3}f_F(k^0)
\int_0^\infty ds\int_{-1}^1du\left({\pi\over is}\right)^{1\over 2}{eB\over \sin z}
e^{-isu m^2}
\nonumber \\
&&\times
e^{-is\Lambda^2}
\left[{e^{-is(1-u)M^2}-e^{-is(1-u)\xi M^2}}\right]
\biggl\{(p_0-k_0)^2e^{-i\sigma_3 z u}\left[-\gamma^0k_0+(1-u)\gamma^3p_3\right]
\nonumber \\
&&+2k^0(k^0-p^0)\left[ue^{-i\sigma_3 z u}\gamma^3p_3+
{\sin z u\cos z\over \sin z}\not\!p_\bot\right]
+{i\over 2s}2e^{-i\sigma_3 z u}\left[\gamma^0(2p_0-k_0)-(1-u)\gamma^3p_3\right]
\nonumber \\
&&-{ieB\over \sin z}e^{i\sigma_3 z(1-u)}\Bigl[-\gamma^0k_0-(1-u)\gamma^3p_3\Bigr]
-u^2p_3^2 e^{-i\sigma_3 zu}\gamma^0k_0
+i{u\over s}e^{-i\sigma_3 zu}\gamma^3p_3
\nonumber \\
&&+{\sin^2 zu\over \sin^2 z}p_\bot^2
e^{i\sigma_3z(2-u)}\Bigl[-\gamma^0k_0+(1-u)\gamma^3p_3\Bigr]
\nonumber \\
&&+\left[{2\sin z u\sin z(1-u)\over \sin^2 z}p_\bot^2+{2ieB\over \sin z}
\right]e^{i\sigma_3 z(1-u)}\left[\gamma^0(p_0-k_0)+u\gamma^3p_3\right]
\nonumber \\
&&+2i\sin zu[(p_0-k_0)-up_0]\not\!p_\bot p_3+2u(1-u)
e^{-i\sigma_3 zu}p_3^2\left[\gamma^0(p_0-k_0)+{u\over 2}\gamma^3p_3\right]
\nonumber \\
&&+\left[2u(1-u){\sin zu\cos z\over \sin z}p_3^2-u^2
{\sin z(1-u)\over \sin z}p_3^2+{\sin^2 zu\sin z(1-u)\over \sin^3 z}p_\bot^2\right]
\not\!p_\bot
\nonumber \\
&&\left.+{i\not\!p_\bot\over \sin z}\left[{\sin z(1-u)\over 2s}+
{\sin zu\cos z\over s}+{2eB\sin zu\over \sin z}-ik_0^2\sin z(1-u)\right]
\right\}\gamma_L
\label{2_25}
\end{eqnarray}
\begin{eqnarray}
&&\Sigma_{bos}^B (p,\xi)
=\Sigma_{bos}^B\left(p,\xi\!\! =1 \!\right)+
{g^2\over 4M^2}\gamma_R\int_{-\infty}^{\infty}{dk^0\over (2\pi)^3}f_B(k^0)
\int_0^\infty ds\int_{-1}^1du\left({\pi\over is}\right)^{1\over 2}{eB\over \sin z}
e^{-is(1-u)m^2}
\nonumber \\
&&\times e^{-is\Lambda^2}\left[{e^{-isuM^2}-e^{-isu\xi M^2}}\right]
\Biggl\{k_0^2e^{-i\sigma_3 z(1-u)}\left[\gamma^0(k_0-p_0)+u\gamma^3p_3\right]
\nonumber \\
&&+2k^0(k^0-p^0)\left[(1-u)e^{-i\sigma_3 z(1-u)}\gamma^3p_3+
{\sin z(1-u)\cos z\over \sin z}\not\!p_\bot\right]
\nonumber \\
&&+{i\over 2s}e^{-i\sigma_3 z(1-u)}\left[\gamma^0(k_0+p_0)-
u\gamma^3p_3\right]-{ieB\over \sin z}e^{i\sigma_3 zu}\left[\gamma^0(k_0-p_0)-
u\gamma^3p_3\right]
\nonumber \\
&&+(1-u)^2p_3^2 e^{-i\sigma_3 z(1-u)}\gamma^0(k_0-p_0)+
{\sin^2 z(1-u)\over \sin^2 z}p_\bot^2e^{i\sigma_3z(1+u)}\left[\gamma^0(k_0-p_0)+
u\gamma^3p_3\right]
\nonumber \\
&&+\left[{2\sin z(1-u)\sin zu\over \sin^2 z}p_\bot^2+
{2ieB\over \sin z}\right] e^{i\sigma_3 zu}\left[\gamma^0k_0+(1-u)\gamma^3p_3\right]
+i{1-u\over s}e^{-i\sigma_3 z(1-u)}\gamma^3p_3
\nonumber \\
&&+2i\sin z(1-u)[k_0-(1-u)p_0]\not\!p_\bot p_3+
2u(1-u)e^{-i\sigma_3 z(1-u)}p_3^2\left[\gamma^0k_0+{1-u\over 2}
\gamma^3p_3\right]
\nonumber \\
&&+\left[2u(1-u){\sin z(1-u)\cos z\over \sin z}p_3^2-(1-u)^2
{\sin zu\over \sin z}p_3^2+{\sin^2 z(1-u)\sin zu\over \sin^3 z}
p_\bot^2\right]\not\!p_\bot
\nonumber \\
&&+\left.{i\not\!p_\bot\over \sin z}\left[{\sin zu\over 2s}+
{\sin z(1-u)\cos z\over s}+{2eB\sin z(1-u)\over \sin z}-
ik_0^2\sin zu\right]
\right\}\gamma_L
\label{2_26}
\end{eqnarray}
where 
\begin{equation}
\Lambda^2=\left[-k_0^2-(1-u)p_0^2+2(1-u)k_0p_0+
(1-u)up_3^2+{\sin  z(1-u)\sin zu\over z\sin z}p_\bot^2\right]
\label{2_27}
\end{equation}
and $z=eBs$. The two Eqs. (\ref{2_25}) and (\ref{2_26})  yield the exact
non-local thermal self-energy of a neutrino, for arbitrary magnetic field,
temperature (as long as it is below the critical temperature of the $SU(2)
\times U(1)$ model), density, neutrino four momentum, and gauge parameter.
This is one of the main results of this paper.
%%%%%%%%%%%%%%%%%%%%%%%%%%%%%%%%%%%%%%%%%%%%%%%%%%%%%%
\section{ $CP$-symmetric medium}
In a medium that is $CP$-symmetric, with a magnetic field $B\ll M^2/e$
and at a temperature $T\ll M$, 
the tadpole diagram vanishes, the scalar diagram and
the contribution of thermal bosons are negligible
and only the bubble diagram with a $W$ and thermal electron
contributes to the self-energy. Under these conditions we write
$\Sigma_{bos}^F (p,\xi)$ as
\begin{eqnarray}
&&\Sigma_{bos}^F(p,\xi)=
\Sigma_{bos}^F(p,\xi\!\! =1 \!)+
{g^2\over M^4}\left(1-{1\over \xi}\right)
\int_{-\infty}^{+\infty} {dk\over
(2\pi)^4}f_F(k)\int_{-\infty}^{+\infty}ds
e^{-[is(m^2-k^2)+|s|\epsilon]}
\nonumber \\
&&\times
\left({\pi\over is}\right)^{3/2}
\gamma_R \left[\left(k^2+{i\over 2s}\right)
{eBs \over \tan eBs}\not\!p+{i\over s}\left({eBs \over \sin eBs}
\right)^2\not\!p-ieBs\left(k^2+{i\over 2s}\right)
\sigma_3\not\!p_\|
\right]
\gamma_L ,
\label{3_01}
\end{eqnarray}
notice that, in order to obtain Eq. (\ref{3_01}), we use
an expansion of Eq. (\ref{2_02} for 
the $W$-propagator in powers of the energy-momentum
transfer $q$ and the magnetic field $B$,
\begin{equation}
G^{\mu \nu}(q)={g^{\mu \nu}\over M^2}\left(1-{q^2\over M^2}\right)-2i{eF^{\mu \nu}
\over M^4}
+\left(1-{1\over \xi}\right)\left({q^\mu q^\nu \over M^4}+{i\over 2}{eF^{\mu \nu}
\over M^4}\right)
+{\mathcal{O}}\left({1\over M^6}
\right)\,.
\label{3_02}
\end{equation}
After the $s$-integration and suitable change of variable, we find
\begin{equation}
\Sigma_{bos}^F(p,\xi)=
\Sigma_{bos}^F(p,\xi\!\! =1 \!)+
{g^2m^2\over 2M^4}\left(1-{1\over \xi}\right){eB\over 2\pi^2}
\int_0^{\infty}\!\!dk_z
\gamma_R\!\left[\sum_{n,\lambda}{f_F(E_{n,\lambda,k_z})
\over E_{n,\lambda,k_z}}\not\!p-{f_F(E_{0,1,k_z})
\over E_{0,1,k_z}}\sigma_3\not\!p_\|
\right]\!\gamma_L
\label{3_03}
\end{equation}
where $E_{n,\lambda,k_z}$ are the energies of the Landau levels of thermal
electrons given in Eq. (\ref{2_18}) and $E_{0,\lambda,k_z}$ is the energy 
of the lowest Landau level.

Using a manifestly covariant notation we write $\Sigma_{bos}^F(p,\xi\!\! =1 \!)$
\cite{erdas2} as
\begin{equation}
\Sigma_{bos}^F(p,\xi=1)=\gamma_R\!\left[
a\not\!p+b\not\!u+c\not \!\!B
\right]\!\gamma_L
\label{3_04}
\end{equation}
where $u^\mu$ is the four velocity of the medium ($\not\!u=-\gamma^0$
in the reference frame of the medium), and
\begin{equation}
a={1\over 2}{g^2\over M^4}{eB\over 2\pi^2}\int^{\infty}_0\!\!{dk_z}
\sum_{n=0}^\infty\sum_{\lambda=\pm 1}
{f_F(E_{n,\lambda,k_z})\over E_{n,\lambda,k_z}}
(E^2_{n,\lambda,k_z}-m^2-k^2_z)
\label{3_05}
\end{equation}
\begin{equation}
b={g^2\over M^4}{eB\over 2\pi^2}\int^{\infty}_0\!\!{dk_z}
\left[{f_F(E_{0,1,k_z})\over E_{0,1,k_z}}
k^2_z{\vec p}\cdot {\hat B}-{1\over 2}
\sum_{n=0}^\infty\sum_{\lambda=\pm 1}
{f_F(E_{n,\lambda,k_z})\over E_{n,\lambda,k_z}}
(3E^2_{n,\lambda,k_z}-m^2-k^2_z)E\right]
\label{3_06}
\end{equation}
\begin{equation}
c=-{g^2\over M^4}{e\over 2\pi^2}\int^{\infty}_0\!\!{dk_z}
\left[f_F(E_{0,1,k_z})
E_{0,1,k_z}E+{1\over 2}
\sum_{n=0}^\infty\sum_{\lambda=\pm 1}
{f_F(E_{n,\lambda,k_z})\over E_{n,\lambda,k_z}}
(E^2_{n,\lambda,k_z}-m^2-3k^2_z){\vec p}\cdot {\hat B}\right]
\label{3_07}
\end{equation}
where the neutrino four momentum is $p^\mu=(E,{\vec p})$ and ${\hat B}$ is
a unit vector in the direction of the magnetic field.
Now we can write the neutrino self-energy in arbitrary gauge as
\begin{equation}
\Sigma(p,\xi)=\gamma_R\!\left[
(a+a_0+a_\xi)\not\!p+(b+b_0+b_\xi)\not\!u+(c+c_\xi)\not \!\!B
\right]\!\gamma_L
\label{3_08}
\end{equation}
where $a_0$ and $b_0$ represent the contribution of the bubble diagram with
$Z$-boson and neutrino in the loop,
which is independent of the magnetic field and of the gauge parameter up to order 
$1/M^4_Z$, and $a_\xi, b_\xi, c_\xi$
are easily obtained from Eq. (\ref{3_03}) using $\gamma_R \sigma_3\not\!p_\|=
\gamma_R(\gamma^3E-\gamma^0 p^3)$
\begin{equation}
a_\xi={1\over 2}{g^2m^2\over M^4}\left(1-{1\over \xi}\right){eB\over 2\pi^2}
\int_0^{\infty}\!\!dk_z
\sum_{n,\lambda}{f_F(E_{n,\lambda,k_z})
\over E_{n,\lambda,k_z}}
\label{3_09}
\end{equation}
\begin{equation}
b_\xi=-{1\over 2}{g^2m^2\over M^4}\left(1-{1\over \xi}\right)
{e{\vec p}\cdot {\vec B}\over 2\pi^2}
\int_0^{\infty}\!\!dk_z
{f_F(E_{0,1,k_z})\over E_{0,1,k_z}}
\label{3_10}
\end{equation}
\begin{equation}
c_\xi=-{1\over 2}{g^2m^2\over M^4}\left(1-{1\over \xi}\right){e\over 2\pi^2}
\int_0^{\infty}\!\!dk_z
{f_F(E_{0,1,k_z})\over E_{0,1,k_z}}E\,\,.
\label{3_11}
\end{equation}

The dispersion relation for neutrino in a $CP$-symmetric medium with magnetic field
is obtained by setting
\begin{equation}
\left[\not\!p+
(a+a_0+a_\xi)\not\!p+(b+b_0+b_\xi)\not\!u+(c+c_\xi)\not \!\!B
\right]^2=0
\label{3_12}
\end{equation}
which gives
\begin{eqnarray}
&&(1+a+a_0+a_\xi)^2(E^2-{\vec p}^2)+(b+b_0+b_\xi)^2-(c+c_\xi)^2B^2
\nonumber \\
&&+2(1+a+a_0+a_\xi)(b+b_0+b_\xi)E-2(1+a+a_0+a_\xi)(c+c_\xi)
{\vec p}\cdot {\vec B}=0.
\label{3_13}
\end{eqnarray}
Using a relationship between the $b_\xi$ and $c_\xi$ coefficients
\begin{equation}
b_\xi={{\vec p}\cdot {\vec B}\over E}c_\xi
\label{3_14}
\end{equation}
and neglecting terms that are proportional to $g^4$, Eq. (\ref{3_13})
becomes
\begin{equation}
(1+2a+2a_0+2a_\xi)(E^2-{\vec p}^2)
+2(b+b_0)E-2c{\vec p}\cdot {\vec B}=0
\label{3_15}
\end{equation}
and, after taking the square root, we obtain
\begin{equation}
(1+a+a_0+a_\xi)E+b+b_0
=(1+a+a_0+a_\xi)|{\vec p}|+c{\vec p}\cdot {\vec B}.
\label{3_16}
\end{equation}
finally, dividing by $(1+a+a_0+a_\xi)$ and neglecting terms of order $g^4$,
we obtain the dispersion relation
\begin{equation}
E=|{\vec p}|-b-b_0+c{\hat p}\cdot {\vec B}
\label{3_17}
\end{equation}
where ${\hat p}$ is a unit vector in the direction of the neutrino momentum.
This dispersion relation is the same that was obtained in Ref. \cite{erdas2} 
and does not depend on either $a_\xi$, $b_\xi$ or $c_\xi$ and thus is independent
of the gauge parameter up to terms of order $(g^2/M^4)$. Therefore the lack of
terms containing $a$, $a_0$ and $a_\xi$ in the dispersion relation tells us
that any term in the thermal self-energy that is proportional to $\not \!\!p$
will not contribute to the dispersion relation. The relationship of Eq. 
(\ref{3_14}) between the $b_\xi$ and $c_\xi$ terms, which in turn leads to a cancellation
of the two terms in the dispersion relation, is due to the fact that both terms come from
the $\sigma_3\not\!p_\|$ term of the thermal self-energy, and therefore 
any term in the self-energy that is proportional to $\sigma_3\not\!p_\|$
will not contribute to the dispersion relation to leading order. 
We would like to stress the fact
this result is general, since the only assumption on the $a$, $b$ and $c$ terms 
of eq. (\ref{3_08}) is that they all contribute to leading order to the self-energy.

The effective potential for neutrinos in a $CP$-symmetric
plasma with magnetic field is easily obtained from the dispersion relation of
Eq. (\ref{3_17})
\begin{equation}
V_{eff}=-b-b_0+c{\hat p}\cdot {\vec B}
\label{3_18}
\end{equation}
and does not depend on $\xi$.

Next we consider the case of a magnetized $CP$-symmetric medium 
at higher temperature, where the contribution of thermal bosons
to the self energy is relevant. The magnetic field is still
$B\ll M^2/e$. In this case, we obtain the following expressions
for $\Sigma_{bos}^F(p,\xi)$ and $\Sigma_{bos}^B(p,\xi)$
\begin{eqnarray}
&&\Sigma_{bos}^B(p,\xi)=\Sigma_{bos}^B(p,\xi\!\! =1 \!)
-i{g^2\over 4}{eB\over M^2}
\int {d^3 {\vec k}\over (2\pi)^3}
\int_0^\infty \!\!\!ds \,e^{-ism^2} e^{-is({\vec p}-{\vec k})^2}
\gamma_R\Biggl[\Bigl\{\bigl[is (\gamma^0\Omega_k
-{\vec \gamma}\cdot {\vec k})
\nonumber \\
&&
\times\sigma_3(\not\!p_\parallel+\gamma^0\Omega_k 
- \gamma^3  k^3)
(\gamma^0\Omega_k-{\vec \gamma}\cdot {\vec k}) 
+\sigma_3(\not\!p_\parallel+\gamma^0\Omega_k- \gamma^3 k^3)
\bigr]e^{is(p^0-\Omega_k)^2}
\nonumber \\
&&
+(\Omega_k\rightarrow-\Omega_k)
\Bigr\}{f_B(\Omega_k)\over \Omega_k}
+{\partial\over\partial M^2}\biggl(
\Bigl\{\bigl[\not\!k_\perp \sigma_3(\not\!p 
+\gamma^0\Omega_k-{\vec \gamma}\cdot {\vec k})
(\gamma^0\Omega_k-{\vec \gamma}\cdot {\vec k}) 
-(\gamma^0\Omega_k-{\vec \gamma}\cdot {\vec k}) 
\nonumber \\
&&
\times
(\not\!p +\gamma^0\Omega_k-{\vec \gamma}\cdot {\vec k}) \not\!k_\perp 
\sigma_3\bigr]e^{is(p^0-\Omega_k)^2}
+(\Omega_k\rightarrow-\Omega_k)
\Bigr\}{f_B(\Omega_k)\over \Omega_k}\biggr)-
(M^2\rightarrow \xi M^2)\Biggr]\gamma_L
\label{3_19}
\end{eqnarray}
\begin{eqnarray}
&&\Sigma_{bos}^F(p,\xi)=\Sigma_{bos}^F(p,\xi\!\! =1 \!)+i{g^2\over 4}{eB\over M^2}
\int {d^3 {\vec k}\over (2\pi)^3}
\int_0^\infty \!\!\!ds \,(e^{-isM^2}-e^{-is\xi M^2} ) 
e^{-is({\vec p}-{\vec k})^2}
\gamma_R\Biggl(
\biggl\{\Bigl[
\sigma_3(-\gamma^0\omega_k
\nonumber \\
&&
+\gamma^3 k^3)
-is(\not\!p_\perp-\not\!k_\perp) \sigma_3
(\gamma^0\omega_k-{\vec \gamma}\cdot {\vec k})
(\not\!p 
+\gamma^0\omega_k-{\vec \gamma}\cdot {\vec k})
+is(\not\!p +\gamma^0\omega_k-{\vec \gamma}\cdot {\vec k})
\nonumber \\&&
\times
(\gamma^0\omega_k-{\vec \gamma}\cdot {\vec k})
(\not\!p_\perp-\not\!k_\perp) \sigma_3
\Bigr]e^{is(p^0-\omega_k)^2}+(\omega_k\rightarrow-\omega_k)
\biggr\}
{f_F(\omega_k)\over \omega_k}
+{\partial\over\partial m^2}\biggl\{\Bigl[
(\not\!p+\gamma^0\omega_k-{\vec \gamma}\cdot {\vec k}) 
\nonumber \\
&&\times
\sigma_3(\gamma^0\omega_k-\gamma^3 k^3)
(\not\!p +\gamma^0\omega_k-{\vec \gamma}\cdot {\vec k}) 
e^{is(p^0-\omega_k)^2}+(\omega_k\rightarrow-\omega_k)
\Bigr]{f_F(\omega_k)\over \omega_k}
\biggr\}\Biggr)\gamma_L
\label{3_20}
\end{eqnarray}
where $\Sigma_{bos}^F(p,\xi\!\! =1 \!)$ and $\Sigma_{bos}^F(p,\xi\!\! =1 \!)$
have been calculated in Ref. \cite{erdas2},
\begin{equation}
\Omega_k=\sqrt{M^2+k^2}
\label{3_21}
\end{equation}
\begin{equation}
\omega_k=\sqrt{m^2+k^2},
\label{3_22}
\end{equation}
and the term $+(\Omega_k\rightarrow-\Omega_k)$ inside the braces in Eq. 
(\ref{3_19}) means that we need to take whatever expression is
contained in the braces, change $\Omega_k$ into
$-\Omega_k$, then add this to the original expression. 
The same meaning have $M^2\rightarrow \xi M^2$ 
and $\omega_k\rightarrow-\omega_k$ that appear later. Notice that while
$M^2$ does not appear explicitly in Eq. (\ref{3_19}), it is contained
inside $\Omega_k$, and thus the partial derivative is meaningful. 
The same is true for the derivative with respect to $m^2$ in Eq. (\ref{3_20}).
After working on the algebra of the gamma matrices, using the following
identities
\begin{equation}
\int d^3 {\vec k}\,e^{2is{\vec p}\cdot{\vec k}}\,k^i={p^i\over |{\vec p}|^2}
\int d^3 {\vec k}\,e^{2is{\vec p}\cdot{\vec k}}\,{\vec p}\cdot{\vec k}
\label{3_23}
\end{equation}
\begin{equation}
\int d^3 {\vec k}\,e^{2is{\vec p}\cdot{\vec k}}\,k^ik^j=
\int d^3 {\vec k}\,e^{2is{\vec p}\cdot{\vec k}}\,{\vec p}\cdot{\vec k}
\left[\left({\delta^{ij}\over |{\vec p}|^2}-{p^ip^j\over |{\vec p}|^4}\right){1\over 2is}
+{p^ip^j\over |{\vec p}|^4}{\vec p}\cdot{\vec k}\right]
\label{3_24}
\end{equation}
and integrating over the parameter $s$, we obtain
\begin{eqnarray}
&&\Sigma_{bos}^B(p,\xi)=\Sigma_{bos}^B(p,\xi\!\! =1 \!)-{g^2\over 4}{eB\over M^2}
\int {d^3 {\vec k}\over (2\pi)^3}
\gamma_R\Biggl[
2{{\vec k}\cdot{\vec p}\over |{\vec p}|^2}{\partial\over\partial M^2}\biggl(\Bigl\{
-(\ln A_++\ln A_-)
\sigma_3\not\!p_\parallel
\nonumber \\
&&
+\bigl[\Omega_k(A_-^{-1}-A_+^{-1})
+{{\vec k}\cdot{\vec p}\over |{\vec p}|^2}
(A_-^{-1}+A_+^{-1})p^0
+{1\over 2|{\vec p}|^2}(\ln A_++\ln A_-)p^0\bigr]
(p^0\sigma_3\not\! p_\parallel-p^3\not\!p)
\Bigr\}{f_B(\Omega_k)\over \Omega_k}\biggr)
\nonumber \\
&&
+\biggl\{\Bigl[A_-^{-1}+A_+^{-1}
-(M^2-2\Omega^2_k)(A_-^{-2}+A_+^{-2})\Bigr]
\sigma_3\not\!p_\parallel
-m^2{{\vec k}\cdot{\vec p}\over |{\vec p}|^2}(A_-^{-2}+A_+^{-2})\gamma^0p^3
\nonumber \\
&&
-m^2\Omega_k(A_-^{-2}-A_+^{-2})\gamma^3 
-2{{\vec k}\cdot{\vec p}\over |{\vec p}|^2}
\Bigl[{{\vec k}\cdot{\vec p}\over |{\vec p}|^2}(A_-^{-2}+A_+^{-2})
p^0p^3 \not\!p 
+\Omega_k(A_+^{-2}-A_-^{-2})
(p^3\not\!p +p^0\sigma_3\not\! p_\parallel)
\nonumber \\
&&
+{1\over 2|{\vec p}|^2}(A_+^{-1}+A_-^{-1})p^0
(p^3\not\!p -p^0\sigma_3\not\! p_\parallel)\Bigr]
\biggr\}{f_B(\Omega_k)\over \Omega_k}-
(M^2\rightarrow \xi M^2)\Biggr]\gamma_L
\label{3_25}
\end{eqnarray}
\begin{eqnarray}
&&\Sigma_{bos}^F(p,\xi)=\Sigma_{bos}^F(p,\xi\!\! =1 \!)+{g^2\over 4}{eB\over M^2}
\int {d^3 {\vec k}\over (2\pi)^3}
\gamma_R\Biggl[
m^2{\partial\over\partial m^2}\biggl\{\Bigl[
\omega_k (B_+^{-1}-B_-^{-1})\gamma^3
\nonumber \\
&&
- {{\vec k}\cdot{\vec p}\over |{\vec p}|^2}(B_+^{-1}+B_-^{-1})\gamma^0p^3
\Bigr]{f_F(\omega_k)\over \omega_k}
\biggr\}
+{\partial\over\partial m^2}\Biggl(\biggl\{(B_+^{-1}+B_-^{-1})\Bigl[2
\omega^2_k \sigma_3\not\!p_\parallel
-2\bigl({{\vec k}\cdot{\vec p}\over |{\vec p}|^2}\bigr)^2
p^0p^3\not\!p \Bigr]
\nonumber \\
&&
+2{{\vec k}\cdot{\vec p}\over |{\vec p}|^2}\omega_k
(B_+^{-1}-B_-^{-1})(
p^3\not\!p +p^0\sigma_3\not\!p_\parallel)
-{{\vec k}\cdot{\vec p}\over 2|{\vec p}|^4}(\ln B_+ +\ln B_-)p^0(p^3\not\! p
-p^0\sigma_3\not\! p_\parallel)
\biggr\}{f_F(\omega_k)\over \omega_k}
\Biggr)
\nonumber \\
&&
+\biggl\{2\bigl({{\vec k}\cdot{\vec p}\over |{\vec p}|^2}-1\bigr)
\Bigl[\omega_k(B_+^{-2}-B_-^{-2})-(B_+^{-2}+B_-^{-2})p^0\Bigr]
(p^3\not\! p-p^0\sigma_3\not\! p_\parallel)
+{{\vec k}\cdot{\vec p}\over |{\vec p}|^4}(B_+^{-1}+B_-^{-1})
\nonumber \\
&&
\times
p^0(p^3\not\!p 
-p^0\sigma_3\not\! p_\parallel)
-2{{\vec k}\cdot{\vec p}\over 
|{\vec p}|^2}(B_+^{-1}+B_-^{-1})
\sigma_3\not\!p_\parallel
\biggr\}{f_F(\omega_k)\over \omega_k}-
(M^2\rightarrow \xi M^2)\Biggr]\gamma_L
\label{3_26}
\end{eqnarray}
where we neglected terms that are proportional to the square of the neutrino
four momentum $p^2_\mu$, and we define
\begin{equation}
A_\pm=m^2-M^2-2{\vec k}\cdot {\vec p}\pm 2p^0\Omega_k
\label{3_27}
\end{equation}
\begin{equation}
B_\pm=M^2-m^2-2{\vec k}\cdot {\vec p}\pm 2p^0\omega_k.
\label{3_28}
\end{equation}
When the temperature is such that thermal boson contributions become important, 
the electron mass can certainly be neglected, and therefore we can ignore terms
in Eqs. (\ref{3_25}) and (\ref{3_26}) that are proportional to $m^2$. All the terms
that are left in the thermal self-energy are now proportional to either $\not\! p$ or
$\sigma_3\not\! p_\parallel$ and therefore, as we showed earlier in this section,
they do not contribute to leading order to the dispersion relation, which is independent
of the gauge parameter.
%%%%%%%%%%%%%%%%%%%%%%%%%%%%%%%%%%%%%%%%%%%%%%%%%%%%%%
\section{Conclusions}
We have studied the dispersion of massless Dirac neutrinos in a magnetized 
plasma. We computed the real part of the
neutrino self-energy in a magnetized medium in an arbitrary
gauge. Our result is exact to all orders in the magnetic field, and valid for all
temperatures below the critical temperature of the electroweak model and all values
of electron density. Then we consider a charge-symmetric plasma at temperature
$T\ll M$ and obtain the $g^2/M^4$ term of the self-energy in arbitrary gauge 
to all orders in the magnetic field. We proceed to compute the dispersion
relation under this conditions and show that, while the self energy 
depends on $\xi$, the dispersion relation is independent of the 
gauge parameter to leading order. We use the dispersion relation
to calculate the effective potential for neutrinos in a charge-symmetric
magnetized medium at temperature $T\ll M$. These matter conditions might
occur in the core of a supernova or in the early Universe, and therefore
it is important to understand the propagation of neutrinos in such
environment. Last we consider the case of a $CP$-symmetric medium
at very high temperature ($T\sim M$), we obtain the self-energy
and the dispersion relation, which is shown to be independent of $\xi$.
%%%%%%%%%%%%%%%%%%%%%%%%%%%%%%%%%%%%%%%%%%%%%%%%%%%%%%
\acknowledgements
A. Erdas wishes to thank Gordon Feldman for helpful discussions
and the High Energy Theory Group of 
the Johns Hopkins University for the hospitality extended to him
during his several visits.
This work is supported in part by MURST (Ministero Universit\`a e 
Ricerca Scientifica).
%%%%%%%%%%%%%%%%%%%%%%%%%%%%%%%%%%%%%%%%%%%%%%%%%%%%%%%%%%%%%%%%%%%%%

%%%%%%%%%%%%%%%%%%%%%%%%%%%%%%%%%%%%%%%%%%%%%%%%%%%%%%%%%%%%%%%%%%%%%
\begin{figure}[htbp]
\begin{picture} (389,200)(0,0)
\ArrowLine (193,114)(221,114)
\ArrowLine (221,114)(278,114)
\ArrowLine (278,114)(307,114)
\PhotonArc(250,114)(28,0,180){5}{9}
\Text(245,100)[l]{(a)}

\ArrowLine(103,28)(131,28)
\ArrowLine(131,28)(189,28)
\ArrowLine(189,28)(217,28)
\DashArrowArc(160,28)(28,0,180){6}
\Text(155,14)[l]{(b)}

\ArrowLine(274,28)(331,28)
\ArrowLine(331,28)(389,28)
\Photon(331,28)(331,57){5}{5}
\BCirc(331,71){14}
\Text(326,14)[l]{(c)} 
\end{picture}
\caption{
The 3 diagrams relevant to the one-loop self-energy calculation; 
(a) the bubble diagram with $W$-boson, (b) the bubble diagram 
with scalar and (c) the tadpole.}
\end{figure}
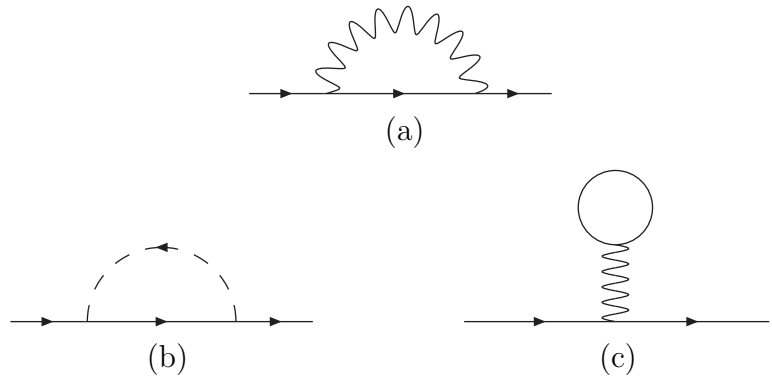
%%%%%%%%%%%%%%%%%%%%%%%%%%%%%%%%%%%%%%%%%%%%%%%%%%%%%%%%%%%%%%%%%%%%%
\end{document}